# Automatic instantiation of abstract tests on specific configurations for large critical control systems

Francesco Flammini[1,2], Nicola Mazzocca[2], Antonio Orazzo[1]

[1] ANSALDO STS - Ansaldo Segnalamento Ferroviario S.p.A.
Via Nuova delle Brecce 260, Naples, Italy
{flammini.francesco, orazzo.antonio}@asf.ansaldo.it
[2] Università di Napoli "Federico II"
Dipartimento di Informatica e Sistemistica
Via Claudio 21, Naples, Italy
{frflammi, nicola.mazzocca}@unina.it

**Abstract.** Computer-based control systems have grown in size, complexity, distribution and criticality. In this paper a methodology is presented to perform an "abstract testing" of such large control systems in an efficient way: an abstract test is specified directly from system functional requirements and has to be instantiated in more test runs to cover a specific configuration, comprising any number of control entities (sensors, actuators and logic processes). Such a process is usually performed by hand for each installation of the control system, requiring a considerable time effort and being an error prone verification activity. To automate a safe passage from abstract tests, related to the so called generic software application, to any specific installation, an algorithm is provided, starting from a reference architecture and a state-based behavioural model of the control software. The presented approach has been applied to a railway interlocking system, demonstrating its feasibility and effectiveness in several years of testing experience.

## 1. Introduction

Computer-based systems used for industrial control applications feature a high number of requirements that make them complex, heterogeneous and highly distributed. When they are also required to be large and safety-critical, system testing, requiring extensive code and configuration coverage, becomes a difficult and time consuming process, which would be almost infeasible without advanced test specification and execution methodologies and tools.

This paper addresses some important issues related to abstract testing of large mission/safety critical control systems. Abstract testing can be defined as a configuration independent and auto-instantiating approach to system testing of large computer-based control installations. In other words, it consists in having an abstract test specification, written without referring to any specific system installation, and a mechanism to automatically detect the specific configuration of the control system and instantiate accordingly the abstract test-suite into test-cases to be physically executed on the system under verification. The configuration data depends on the type and number of devices to be used, which in turn is usually installation specific, while the control algorithm is configuration independent in most cases[1]. This means that control actions performed by the actuators depend on device classes and subsets related to the specific installation and on their interrelationships; however, such dependency does not impact on the generality of system functional requirements and of the corresponding test specification. A large class of complex computer-based control systems are also configuration-critical, that is to say they need a thorough (static and/or dynamic) verification of their configuration database. While a static verification can be performed "off-line" (by hand or by means of specific tools), a dynamic verification requires the execution of tests on the target system. In order to achieve a dynamic coverage of the configuration data, more executions of the same abstract test are needed, with the aim of involving more control entities (theoretically, all the combinations should be tested). This time consuming activity can be automated by using the abstract testing approach presented in this paper.

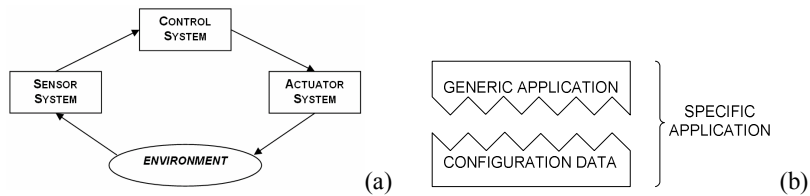

Figure 1. A control system (a) and the integration of control software (b).

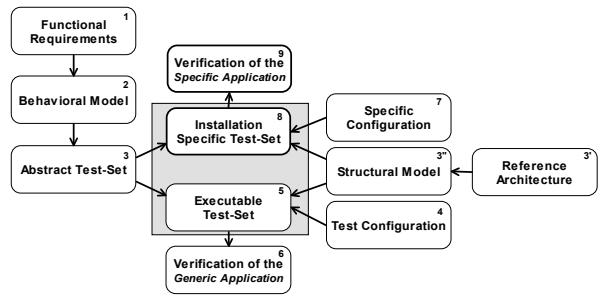

Figure 2. System-level functional V&V activities and paper contribution (shaded box).

---

[1] This assumption holds well for mature domains, where designs can be reused and configured to many different installations; however, the approach may be more difficult to apply to new application areas, which still need to be completely understood in terms of abstract requirements of system logic.

Figure 1a gives an at a glance general representation of a control system, featuring sensor and actuator subsystems in order to interact with the external environment, while Figure 1b shows the necessary integration between generic control software and specific configuration data, which is the central topic of this paper.

Figure 2 reports a general scheme of system V&V activities for safety-critical system installations, with the numbers indicating the usual temporal sequence of activities (steps 3' and 3'' are absent in traditional approaches). The shaded box in the central part of the figure indicate the target of this paper. The two boxes with the thick border represent two recurrent activities, as they must be repeated for each system installation, for which automation is particularly advantageous. Figure 2 also shows that a structural model (step 3'), derived from a reference architecture (step 3''), is needed to achieve automation; this is also a novel achievement, which will be discussed in detail in section §3.

The approach presented in this paper is based on assumptions which define a reference software architecture of the control system. The approach is then applicable whenever the system under test fits or can be refactored/engineered to fit such an architecture. Another important aspect is related to the reference behavioural model, representing system functional specification by means of a Finite State Machine (FSM): the approach can be applied whenever the dynamic of the control system (possibly featuring an unlimited state space) can be abstracted as a FSM. This is true for most computer-based control systems based on a sequential computation, belonging to the Discrete Event (Dynamic) System (DES or DEDS) category (a large amount of literature is available on this subject, as reported in Section §2).

The abstract testing methodology described in this paper has been applied to a railway interlocking (IXL) system, used for train route and ground signalling management. A traditional approach for the system verification of a railway IXL consists in the following steps:

- Functional testing of the control software and measurement of code coverage;
- Static verification of the configuration data, using proper support tools;
- Acceptance testing, regarding the verification of the most significant railway logic conditions.

While quite effective, such an approach does not guarantee dynamic configuration coverage. In other words, control software is tested dynamically over a little part of a specific configuration, until code coverage is considered satisfactory, but there is no evidence that the integration of generic control algorithms and configuration data is correct for the specific installation. A dynamic black-box testing of the specific installation, ensuring both code and configuration coverage, represents the safest approach, considering that installations can be very different one from each other, thus possibly stimulating the control software with untested combinations of inputs. The approach presented in this paper perfectly suited such a real-world industrial case-study, allowing test engineers to rapidly verify large installations, achieving the required coverage for both control code and configuration data[2].

The remainder of this paper is organized as follows: Section 2 sets the contribution of this paper in the context of related works; Section 3 presents the abstract testing methodology, describing the a reference structural model of computer-based control systems and the working principles of the abstract testing algorithm; Section 4 introduces the main characteristics of a railway interlocking system and describes the application of abstract testing to such a case-study; finally, Section 5 summarizes results and gives some hints about future developments and applications.

## 2. Related works

Abstract testing belongs to the category of model-based testing approaches [3]. In model-based testing, test-cases are derived in whole or in part from a model that describes some (usually functional) aspects of the system under test. While many research works address the problem of the automatic generation of the abstract test-suite starting from a system's (formal) model (see reference [7]), very few deal with the problem of the automatic instantiation of abstract test-suites. For instance, reference [8] addresses the problem of executing abstract tests on distributed software.

Reference models are usually based on notations of the Unified Modelling Language (UML) [9]. UML is a de facto standard in software development (see Model Driven Engineering, [10]), but it can be also used for different purposes, like formal verification [11] and reverse engineering [12]. There exist several research works which deal with the automatic test generation from UML models [13]. The Finite State Machine (FSM) formalism, available in an "extended" version amongst UML diagrams as the State Diagrams (or Statechart) view, has also been widely used as a fundamental behavioural representation for test-generation, both informally [6] and formally [14]. A fundamental paper on model-based testing [22] underlines the key concerns in adopting FSM based methods.

Railway control systems are the ideal candidate for the application of the methodology described in this paper, featuring large, distributed, heterogeneous architectures and varying configurations according to any specific installation. The verification of railway control systems (like the ones compliant to the new ERTMS/ETCS standard [26]) have been studied in a number of research works, using either simulation based [27] or formal approaches [24]. As for any safety-critical system, the verification and validation of critical railway control systems must respect international safety standards, like the CENELEC norms [1]. These norms prescribe a number of thorough V&V activities, including hazard-analysis [28]. The distinction between generic and specific software applications shown in Figure 1b is also provided by the CENELEC standard. However, standards prescribe just general requirements and guidelines, but they do not address the solution of specific verification issues, which is under the responsibility of each supplier. From the point of view of the CENELEC standard, this paper covers the methodological aspects related to the verification of the "Specific Application".

---

[2] A widespread and effective code coverage measure consists in Decision Coverage (or one of its variants, like Decision to Decision Path, DDP) [2], while for the configuration the coverage measure refers to the access to the entries of the configuration database.

## 3. The abstract testing methodology

This section describes the abstract testing methodology, starting from a reference architectural model and then presenting the algorithm used to automate the instantiation of functional tests on specific configurations.

### 3.1 Basic definitions

**Definition 1.** A *Control Logic* is defined as the set of software components which implement system functional requirements in order to realize a given control algorithm.

**Definition 2.** A *State* of the system under test is defined as the set of values which are assigned to all its internal state variables. A reduced representation of system state consists in considering only the variables of interest for the Control Logic.

**Definition 3.** An *Abstract Test* is a non executable test-case in which the reference to the involved control entities is symbolic and possibly related to equivalence classes.

**Definition 4.** An *Input Sequence* is a temporal sequence of stimuli provided to system input gates, including sensing devices and human-machine interfaces.

**Definition 5.** An *Output* is the set of data or control actions which are produced by the system at its output interfaces / actuators in response to a certain Input Sequence.

### 3.2 Reference architectural model

A control system always features:

- a Sensor system, constituted by a variable number of possibly heterogeneous sensors, used to detect inputs from the environment;

- an Actuator system, used to implement the control actions decided by system logic;

- a Control system, which collects inputs and elaborates system outputs according to the desired control function and acting on the actuator subsystem.

These main subsystems and their interrelationships are depicted in the white boxes of Figure 3, according to the notation of UML Class Diagrams.

In this paper it is assumed that computer-based control systems (also known as Real-Time systems) implement the control function by means of discrete control logic, written in a suitable programming language, and a configuration database which is used to map the control logic on a specific installation.

The configuration independent entities of the control software constitute the so called "Generic Application" and are shown in light grey in the class diagram of Figure 3; the implementation specific ones are shown in dark grey, and constitute the "Specific Application".

Processes shown in Figure 3 should be meant as software entities managing specific data and functions, while satisfying to a certain extent the object-oriented design paradigm (e.g. data encapsulation). Widespread best-practices in real-time software engineering employ design approaches based on modularity and abstraction, which emulate object-orientation even using non object-oriented (i.e. legacy or proprietary) programming languages [24]. In such a view, processes are associated to physical or logical entities. All processes feature a data structure and operations, and can be scheduled as independent tasks by a real-time operating system, i.e. at each elaboration cycle:

- Sensor Processes collect and manage data easured by sensors;

- Logic Processes cooperate in order to implement the desired control function by accessing the status of Sensor Processes and issuing commands to Actuator Processes;

- Actuator Processes verify commands' actability and possibly implement them by driving actuators.

**Definition 6.** A *Logic Process* is defined as an entity of the control software with a well defined role and structure but without a direct correspondence with a physical entity (i.e. a sensor or an actuator).

In a typical centralized control system, the number of logic processes is usually less than the number of physical objects, but their complexity is usually higher, as they implement the core of the control algorithm.

Despite its specific representation mechanism, the configuration database must provide the instantiation of objects for Sensor and Actuator classes of Figure 3 and their interrelationships with logical entities. This is achieved by defining:

- The physical entities used in the specific installation, by means of Sensor and Actuator Lists;

- The relationships (type, cardinality, etc.) between Sensor/Actuator Processes and Logic Processes, using proper Association Lists.

For coherence of representation, the aforementioned lists are shown as objects in the class diagram of Figure 3, but obviously they can be represented differently when using a non object-oriented database (e.g. by tables in a relational database).

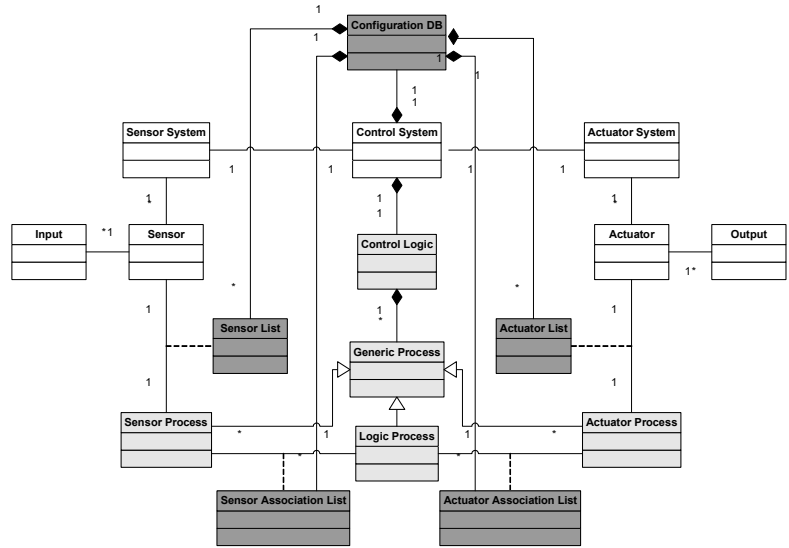

Figure 3. Reference architecture for the computer-based control system.

As already mentioned, the approach presented in this paper is applicable if the software architecture of the system under verification is compliant to the reference model presented above. If the software of the control system has not been yet developed, then the "design for verifiability" philosophy would suggest to make it compliant to the reference architecture for abstract testing. This requires adopting an object-oriented or analogously structured design approach, in which physical and logical entities are mapped on corresponding processes, each one featuring its own attributes and operations. For existing systems which are not compliant to the reference architectural model, it would be necessary to review the design to determine whether it can be re-factored or re-engineered [12]. Although this may involve a good deal of effort, if many configurations are likely then the effort is more than offset by the savings achieved by the multiple application of the abstract testing method.

### 3.3 Abstract testing

Before introducing the abstract testing methodology, some introductory statements are necessary. In this paper, abstract testing is not meant as a functional test specification methodology. Instead, abstract testing allows test engineers to specify configuration independent tests, to automatically instantiate and execute the abstract test-suite on any specific configuration. This allows test engineers to achieve the goal of an extensive coverage of both control code and configuration data in order to verify their correct integration. Abstract testing can not discover configuration errors which do not lead to incorrect system outputs when combined with control code. In fact, the configuration of the system is regarded as an input of the transformation algorithm from abstract to specific tests. Therefore, abstract testing can not detect whether a system is properly configured or not for the specific installation (such verification should be performed separately, and is not in the scope of this work).

The output of the needed abstract testing algorithm is a set of test-cases to be physically executed on the system under test, depending on its specific configuration. In other words:

$$(\text{Abstract Test - Set, Configuration}) \xrightarrow{\text{Tranformation Algorithm}} \text{Physical Tests}$$

<div align="right">Expression 1</div>

Expression 1 means that the algorithm transforms a couple, constituted by an abstract test specification and a specific configuration, into a set of "physical" tests which can be directly executed on the target system (or any simulated version of it [31]).

The cardinality of the transformation is in general "one to many": at least one physical test must be executed for each abstract test, but more of them could be necessary. Nevertheless, a degenerate although possible case consists in a configuration which does not allow executing any physical test for a specified abstract test.

Abstract test specifications are derived from the system functional requirements and are couched in a precise formalism to avoid interpretational ambiguity by the automated algorithm. Functional requirements for the systems of interest are usually specified in natural language, and generally conform to the following template:

> "When system is in state $S_I$ and receives an input I from sensors SEN, then it shall actuate output O using actuators ACT and transit in state $S_O$" where S, I, O, SEN, and ACT are respectively lists, vectors or equivalence classes (determined by particular properties) of states, inputs, outputs, sensors and actuators. Please note that the sensor class SEN is not restricted to be related to the physical sensors of the control system, but can represent any measurement or human interaction device of system interface with the environment (an analogous consideration holds for ACT). In such a way, timing requirements can be managed in the same way of any other requirements. In fact, time can be considered as a variable of interest (whenever required) and mapped both on input conditions to be generated and outputs to be verified. Herein after, with no loss of generality, it will be assumed to be dealing with generic "properties", used to select objects of any class (i.e. S, I, O, SEN, ACT) within requirements, coherently with an abstract specification which should identify entities only according to their properties of interest (i.e. attributes' range of values and relationships with other entities). Usually, informal specification only indicates changes in outputs or output states, assuming the rest remains the same; obviously, this does not influence the generality of the proposed form.

Therefore, a general format for abstract test description (or Test-Case, TC), formalizing the functional requirement, could be the following:

$$(\text{STATE}_I , \text{INPUT}) \rightarrow (\text{OUTPUT, STATE}_O)$$

<div align="right">Expression 2</div>

That it to say, a couple constituted by an output and output state is associated (thus the arrow) to any significant input and input state, as stated by system specification. $\text{STATE}_I$ and $\text{STATE}_O$ represent respectively input and output states. INPUT includes both input values and the involved sensors[3]; similarly, OUTPUT also refers to both actuators and output values. Therefore INPUT = {I, SEN} and OUTPUT = {O, ACT}. Each macro variable in the left part of Expression 2 is a combination of elementary variables, namely "influence variables" [5] (e.g. $STATE_I = (State_{I1}, State_{I2}, …)$), which satisfy a given condition (e.g. $S_I$) and have to be instantiated according to such condition in order to generate an executable test-case (e.g. $State_{I1} = s_{I1}$, $State_{I2} = s_{I2}$, …). Dealing with critical systems, it is assumed to consider any combination of inputs respecting $S_I$ and I, despite of possible redundancies which could be eliminated by defining proper (i.e. safe) reduction criteria to be applied on the test-set [5]. The macro variables on the right of Expression 2, instead, must be checked after test execution in order to verify that their instances satisfy the O and $S_O$ conditions. Note that in general $STATE_I \neq STATE_O$ (intended as sets), that is the state variables to be checked do not have to be the same defined in input state. This leaves test engineers free to define different subsets of interest on input and output states, thus implicitly defining equivalence classes. Finally, note that according to the general structure presented in previous section, system state is exhaustively given by defining the value of all attributes of all its Processes. Of course, such a generalization comprises the simplest cases, in which e.g. the requirement (and thus the test) specifies single input and output instances.

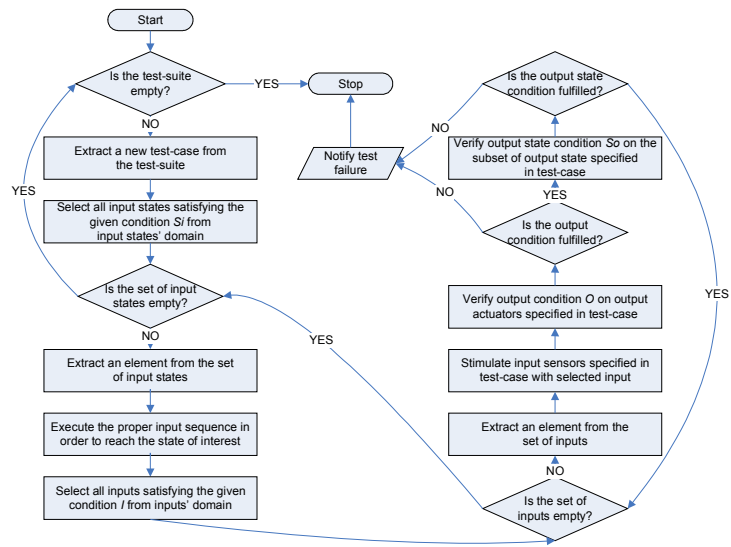

Figure 4. A high level flow-chart of the abstract testing algorithm.

---

[3] Please note the distinction between the physical sensor input (e.g. temperature) and its logical representation (e.g. numerical value). For testing purposes, the control system is often used in simulation environments which are able to inject logical values directly into the system instead of creating real environmental conditions (an obviously more difficult task).

The first two variables, namely $STATE_I$ and INPUT can be collapsed into a single variable, namely INPUT_SEQUENCE. Introducing INPUT_SEQUENCE allows considering the system combinatorial from an input point of view: starting from a well-know INITIAL_STATE of the system, e.g. the one following system boot-strap or initialization, an INPUT_SEQUENCE uniquely determines $STATE_I$ (and possibly also INPUT) for a given test, passing through a number of intermediate system states. This is important, as system functional tests should be always executed starting from a well-known reference system state (typically an "idle" state), and performing all the actions necessary to reach the state of interest (this sequence of steps is often indicated as the "preamble"). Then, system is stimulated with required input, and finally output and output state are checked for correctness with respect to the expected ones (the process is usually automated in simulation environments providing scripting capabilities).

The construction of the preamble requires little effort, indeed, as it can be easily obtained by properly assembling test sequences. In fact, with the exception of the initial reference state, the input state of a test-case (say $TC_X$) is the output state of at least another test-case (say $TC_Y$); formally: $STATE_{I-X} = STATE_{O-Y}$. Of course, the sequence of definition of the test-cases can be advantageously arranged in an order such that:

$$\forall\, TC_X(t)\, \exists\, TC_Y(\bar{t})\,|\,(\bar{t} < t) \wedge (STATE_{I-X} = STATE_{O-Y})$$

Expression 3

In Expression 3, TC (t) represents the test-case defined at time *t*. In this way, the needed preamble is always obtainable by a backward iteration (stopping at the initial state) which selects the sequence of test-case needed to get to the state of interest. This aspect, which can be easily automated, has not been explicated in the algorithm presented in this section, as it represents only a collateral feature.

On the basis of such assumptions, a transformation algorithm for abstract testing can be introduced, with the aim of being (at least partially) automated[4]. The algorithm is provided in two different forms: a classic flow-chart (see Figure 4), which is at a higher abstraction level, less formal but more readable; a more detailed meta-language program (see Figure 5), which uses a sort of (C and SQL)-like pseudo language, featuring a quite self-explaining syntax (variables are shown in Italic font to distinguish them from keywords).

The following further assumptions are necessary in order to simplify the algorithm in its detailed form, without loosing its generality:

- Test-cases are ordered by their input states, so that for any new test-case processed by the algorithm it is possible to find a previously executed (and not failed) test-case whose output state consists of the input one of the new test to be executed: in such a way, the input sequence can be determined backward in a very straightforward manner (the corresponding procedure is omitted in the algorithm for the sake of brevity);

---

[4] Variants of the algorithm are certainly possible, but they should not differ substantially from the one proposed in this paper.

- The complete system state is given by the value assigned to all attributes of all its objects (i.e. $s_1$=attribute$_1$, $s_2$=attribute$_2$, etc.): in such a way, the output state is checked by identifying state variables by the name of the corresponding attributes (assuming them as unique identifiers).

The detailed algorithm written in meta-language is reported in Figure 5. Numbered comments have been added for any significant block of statements (they will be referred to as "steps"), in order to facilitate the understanding of the algorithm. The variables used in the algorithm and their meaning are listed in Table 1 (ordered as they are met in the code).

```
1   /* 0. Scan all abstract test-cases
2   for each TC
3     /* 1. Cycle through all input states of the equivalence class
4     select I_States from Input_States satisfying Si
5     for each STATEi in I_States
6       execute Input_Sequence reaching STATEi
7       /* 2. Select all sensors involved in the test
8       select Input_Sensors from Sensor_List satisfying INPUT->SEN
9       /* 3. Cycle through all sensors to assign their input values
10      for each Sensor in Input_Sensors
11        /* 4. Each sensor is stimulated with an input of the equiv. class
12        select Input_TC from Input satisfying INPUT->I
13        for each INPUTj in Input_TC
14          stimulate Sensor with INPUTj
15      endfor
16      /* When all sensors have been stimulated with proper inputs
17      /* the corresponding output and output state are checked
18      /* 5. Select all actuators involved in the test
19      select Output_Actuators from Actuator_List satisfying OUTPUT->ACT
20      /* 6. Verify that each actuator satisfies its output condition
21      for each Actuator in Output_Actuators
22        check Actuator for condition O
23        if check failed then notify failure
24      endfor
25      /* 7. Select the subset of system state to be checked
26      select all Attributes from STATEo
27      /* 8. Verify that the value of each attribute of control
28      /*    processes satisfies output state condition
29      for each Attrib in Attributes
30        /* 9. Scan through attributes of all processes using association
31        /*    lists to detect the attributes of interest
32        for each Sensor in Input_Sensors
33          select S_Attrib of Sensor where S_Attrib->name=Attrib
34          select Sen_LP in Sensor_Association_List including Attrib
35        endfor
36        for each Actuator in Output_Actuators
37          select A_Attrib of Actuator where A_Attrib->name=Attrib
38          select Act_LP in Actuator_Association_List including Attrib
39        endfor
40        /* 10. Merges selected logic processes in a single list
41        merge Sen_LP and Act_LP to Logic_Processes
42        /* 11. Select attributes of logic processes by their name
43        for each Logic_Process in Logic_Processes
44            select LP_Attrib of Logic_Process where LP_Attrib->name=Attrib
45        endfor
46        /* 12. Merges all selected attributes in a single list
47        merge S_Attrib, A_Attrib, LP_Attrib to Proc_Attrib
48        /* 13. Checks all selected attributes to verify output state
49        check all Proc_Attrib for condition So
50        if check failed then notify failure
51      endfor
52    endfor
53  endfor
54  /* 14. If no fail is notified, test can be considered as "passed"
```

Figure 5. The abstract testing algorithm.

As already mentioned, properties (e.g. I) are generally used in order to extract objects from a given set (e.g. input domain), using proper queries, whose implementation is application specific, depending on the particular representation of the configuration database. The involved Lists are the ones represented in Class Diagram of Figure 3 and already described in previous section. The variables of the algorithm which have not been already defined correspond to lists of objects (or records) obtained by a query (e.g. SELECT Input_Sensors …) on the configuration database, using the input conditions defined by the abstract test.

The cycle numbered as 1 selects a state of the equivalence class defined in the abstract test-case and executes the proper input sequence needed to make the system transit in that state. In such a case, the SELECT query is implicit in the statement, as no system database access is necessary (as aforementioned, each of the input states is generated by extensive combination of elementary states variables defined by test engineers). All following instruction blocks behave in a similar way: they first select a subset of interest of a certain domain by performing a query based on a specified property, and then execute a cycle on the extracted subset.

As aforementioned, in general SEN, ACT, I, O and S could be either lists themselves, and in this case they will contain the identifiers of the entities to be involved (e.g. SEN = {$SEN_1$, $SEN_2$, …}), or properties to be respected by one or more attributes (e.g. SENSOR_TYPE), identifying a class of entities (e.g. SENSOR_TYPE = Temperature_Sensor OR Light_Sensor): both options can be collapsed into the same case of a property based selection, with no loss of generality. In order to get a test specification which is configuration independent, when the selection regards system entities the explicit form should not be used.

Looking at the algorithm, it is evident that for each produced test-case a sensor can be stimulated by a single input at each input state, so there are no input sequences possible at the test-case level (they are only possible at a higher "scenario level"). Such aspect is coherent with the state machine assumption: since the first input can possibly trigger a state transition, following inputs must correspond to different test-cases.

The core of the algorithm consists is located in its second half (steps from 7 to 14), where the output state is checked for correctness. When system is configured on a certain installation, its hardware structure is well known in terms of needed sensors and actuators, while the type and number of logic processes are not directly known. The reason is that logic processes are automatically instantiated according to the hardware configuration, as defined in sensor, actuator and association lists. Therefore, the algorithm is such to access system internal state by scanning the attributes of all control processes, starting form sensor and actuator lists and accessing related logic processes by means of the association lists. In fact, these lists are meant to link system hardware to its software components, both of which are variable from installation to installation. The approach allows moving from external entities (i.e. system interface with the environment) to its control logic (i.e. software processes) by using configuration information. The search process is based on the assumption that the state variables to be checked have the same names of the corresponding attributes. In fact, although object-based programming should avoid attribute duplication, the same state variable can be stored in more homonymous attributes of different classes. A different design option consists in copying at each elaboration cycle the content of attributes constituting the data structure of all objects into a single database: this option is highly advantageous as it avoids attribute duplication (any attribute is a primary key in the database) and simplifies the search process for output state checking (a simple query for each state variable is necessary).

As for algorithm execution, there are two possibilities, which are perfectly equivalent for the purpose of the algorithm:

1) the algorithm for abstract testing is interpreted in real-time, and then statements like "stimulate Sensor with Input" are physically executed on the system under test as the algorithm executes;

2) the algorithm does not directly execute statements, but writes them using a proper syntax on a set of script files to be later executed in an automated testing environment.

One important aspect of the algorithm is that, besides generating test-cases, it also checks output for correctness in an automated way. If the state of the system under verification is not accessible by test engineers, then the part of the algorithm meant to verify output state (steps 7-13) is not applicable. Furthermore, if an automated checking of actuators' output is not possible, steps 5 and 6 are not applicable, and the algorithm only serves as a test-case generator.

## 4. The computer-based railway interlocking case-study

This section describes the application of the abstract testing methodology presented in the previous section to computer-based railway interlocking systems.

### 4.1 Computer-based railway interlocking systems

A railway interlocking system (IXL) is a safety-critical distributed system used to manage train routes and related signals in a station or line section (which is divided into "blocks"). Its development and verification process must respect international safety norms [1]. Modern IXL systems are computer-based and feature a high number of functional requirements, thus making them very complex. The verification and validation (V&V) process of such systems comprises a set of time-consuming activities (hazard-analysis, code inspection, structural testing, etc.), among which functional testing is one of the most important, in terms of both budget and criticality [29]. Moreover, IXL installations are different from each other. Therefore, V&V consists first in verifying the generic software application, and then the specific one. However, while an abstract test-suite can be developed on the basis of the Generic Application (and hence of system functional specification), in order to verify the specific application the former test-suite must be instantiated according to the configuration of the installation under test. Clearly, this need perfectly fits the purpose of the abstract testing approach presented in this paper. Traditionally, such an activity is performed by hand, with evident disadvantages in terms of required effort and correctness of results.

A computer-based IXL is composed of the following entities:

- a safety-critical centralized elaboration unit (indicated with CPU), which is meant to run the control software (processes and configuration);

- a Man Machine Interface (MMI), consisting of a display and a functional keyboard which allows the setting and control of train routes;

- a Communication Computer (CC), used to manage the communication via a Wide Area Network (WAN) with a (distant) central Automation System (AS), also providing remote route management, and possibly adjacent IXL;

- a set of Track Circuits (TC[5]), used to detect if a train is occupying the route;

- a set of Switch Points (SP), used to form train route;

- a set of Light Signals (LS), used to notify to train drivers route status.

The IXL configuration associates each possible route to its related physical control entities: TC, SP and LS.

A slightly simplified architecture of an IXL is reported in Figure 6a.

An IXL is basically used to manage route formation commands coming from a local human operator (using the MMI) or a remote operator (using the AS). When a command is received, the CPU controls its actability by checking the status of all involved entities, either physical (TC, SP and LS) or logical (e.g. block orientation, line out of service, station emergency, etc.). If the route formation command can be safely carried out, then Switch Points are moved accordingly. A route can also be formed in a degraded mode, in which route integrity can not be assured because a check failed on a Track Circuit due to its being occupied, or because a Switch Point is not operational. These degradations reflect on route integrity status, which have to be properly notified by multi-aspect Light Signals. Moreover, the system has also to manage the change of route status when a train passes on it, until the liberation of the route. The state machine associated to a route is quite complex: for the sake of simplicity, it will not be described in detail.

First of all, it is important to distinguish between sensor and actuator entities in an IXL. Clearly, Track Circuits can be considered as Sensors, as they are only used to detect train position. Switch Points and Light Signals are instead Actuators, because they are responsible for system control actions. The interface for route setting and monitoring, finally, can be considered both as a Sensor, as it receives commands, and an Actuator, as it displays outputs; analogously for the WAN interface. Another option, which is equivalent in theory but could be advantageous in practice, is to consider the effects of MMI and WAN interfaces directly on system state: intuitively, moving an input variable into a state variable is always possible and does not necessarily impact on test accuracy.

---

[5] The TC acronym has been formerly used for "Test-Case". In this section it only means "Track Circuit".

Now, a possible IXL software architecture can be defined, using an object oriented design, which will be used as a reference for the abstract testing application. Control processes will be associated to each physical control entity (TC, SP, LS), thus obtaining:

- TC_Process, dealing with Track Circuit status (clear, occupied, broken, etc.);
- SP_Process, managing Swich Point status (straight, reverse, moving, out of control, etc.) and operations (move_straight, move_reverse, etc.);
- LS_Process, aimed at controlling Light Signals' status (green, red, yellow, flashing yellow, etc.)

Furthermore, logic processes have to be defined for each logical control entity, thus obtaining:

- Route_Process, managing route status and control actions;
- Line_Process, managing out of service conditions, temporary speed restrictions, etc.;
- Block_Process, managing logical block orientation;
- LeftIXL_Process (managing data received by left adjacent IXL)
- etc.

Also Sensor type processes, if implemented according to object orientation, shall feature the operations needed to access the status of their attributes.

An IXL will then feature a real-time kernel scheduling the above mentioned processes. In case of the ASF Interlocking system, there exists a Logic Manager used to interpret and schedule processes written in an application specific logic language (see reference [24] for a brief description of its syntax), which allows for a sort of object orientation, even though not being specifically object oriented (see Figure 6b). Moreover, a separate and well defined system configuration will allow customizing of the IXL to each specific installation (e.g. Manchester railway station). In ASF implementation, the configuration database is of a relational type, featuring tables containing "lists of entities" and "lists of linked entities", expressing the interrelationships between logic and physical control processes. Such lists are used to perform queries in order to determine the associations needed for the system to work and which abstract testing is based on (they basically correspond to Sensor/Actuator_Association_Lists referred to in Figure 3). Finally, the output state of the CPU is copied in a "state of entities" database at each elaboration cycle, with attributes being primary keys, thus significantly simplifying output state checking, as explained in previous section.

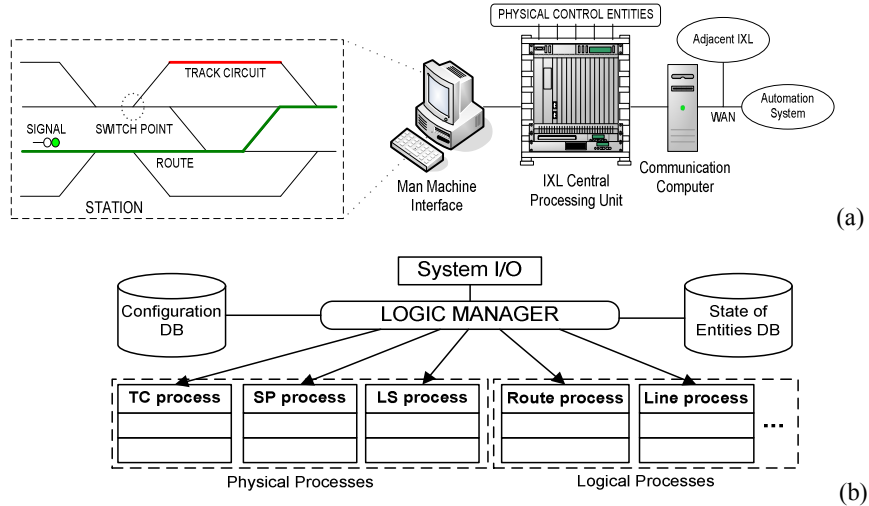

(a)

(b)

Figure 6. An IXL scheme (a) and related control software architecture (b).

### 4.2 Application of the abstract testing methodology

In the previous section, the generic architecture of an IXL and how it can be mapped on the general scheme of Figure 3 have been shown. This mapping being possible, the application of the algorithm in its general form to the IXL case-study is quite straightforward. In order to understand how the algorithm applies to a specific abstract test-case, the following requirement can be considered:

> "When the IXL receives a route formation command, it has to check the following conditions:
>
>> All TC associated to the route must be clean
>>
>> All SP associated to the route must be controlled
>>
>> The LS associated to the route must be controlled
>
> If such conditions are fulfilled, then SPs have to be moved accordingly, route LS shall be set to GREEN and Route Status has to be set to Set_OK"[6].

The requirement can be tested using the following abstract test-case ($r$ represents a generic route):

- Input State:

    Route $r$ TC Status = Clean

    Route $r$ SP Status = Controlled

---

[6] In a real IXL, the correct execution of a switch movement command must be verified before setting the signal to GREEN (in other words, SPs feature additional sensors).

>>>>Route *r* LS Status = Controlled

>>- Input:

>>>>MMI Input: Route *r* Formation Command

>>- Output:

>>>>Route *r* SP Output = Positioned according to *r*

>>>>Route *r* LS Output = GREEN

>>- Output State:

>>>>Route *r* Status = Set_OK

For such an abstract test-case, the algorithm:

- Step 1: Selects all input states fulfilling the *Si* condition specified in the test-case. In other words, all combinations of TC, SP and LS statuses associated to each configured route have to be set to the specified values; for the specific test-case, there exists exactly one combination of such values for each possible route, corresponding to a nominal system state (e.g. the idle state following system start-up);

- Steps 2-4: Stimulates the IXL via the (simulated) MMI with the specific route formation command; as only one command is possible for each route, the *SEN* and *I* conditions will select exactly one sensor and one input for each input state;

- Steps 5-6:

> Checks that all SP are positioned according to the specific route; the actuator selection routine applies to switch points of the specific route: condition *ACT* will read as follows "Actuators SP whose Routes attribute contains *r*" (SP are associated to more than one route); condition *O* will be "Positioned according to *r*";

> Checks that the specific route status is set to GREEN: the check for LS output is analogous to the one described for SP (see point 0 above);

- Steps 7-13: Checks that output status satisfy the *So* condition, that is "Route *r* Status = Set_OK". As attributes are not duplicated into different objects and a "state of entity" dynamic database is available, such a check consists of a simple database access, verifying that "Route_Status_r" (primary key) is set to the value "Set_OK".

Clearly, the algorithm will produce as many physical tests as the number of possible routes which are configured in the IXL. The example is straightforward, corresponding to a nominal test. A possible negative test, related to the previous one, can be generated from the following requirement:

> "[…] If any of the conditions listed above is not fulfilled, then route command must not be accepted".

In such a case and without adopting any reduction criterion on *Si*, there would be a significant increase in the number of generated physical tests. In fact, for each route the combinations of input states impeding the correct formation of the route are numerous, and all of them should be tested.

### 4.3 Results

The application of the abstract testing algorithm serves to automate two important steps in the traditional verification of railway IXL systems:

1. the instantiation of the abstract test-set to a specific installation, in order to make it executable (Generic Application verification stage);

2. the coverage of the so called (railway) "condition table", reporting the expected outputs against all the significant railway conditions that can happen on all possible routes[7] (Specific Application verification stage).

The activity of point (1) usually requires hundreds of person-hours if executed by hand; if automated by means of the proposed abstract testing algorithm, effortless and safe results can be obtained in a few minutes (typically). The very first execution of the test-suite is aimed at checking system control logics, sometimes using a "draft" configuration database. In this case, the target of the coverage measure is the control code. When all the V&V activities of the "Generic Application" are considered satisfactory, the related Safety-Case is issued.

Point (2) involves all the activities aimed at verifying the configuration database of the final installation. The abstract testing algorithm is able to cover all possible conditions on all possible routes. The testing against the condition table has sense, as it cannot be guaranteed a priori that a system output in response to a condition which is valid for one route continues to be valid for a different route. As an example, the decision coverage of the control code would not change by testing conditions on "symmetrical" routes (i.e. different routes which are identical in terms of their physical structure, type of involved entities and railway logic rules), as the code would be exercised in the same way. However, the involved configuration data does change, and its incorrectness or misinterpretation could alter the expected result. The verification against configuration incorrectness is usually performed by means of a static analysis, possibly executed by diversity-based approaches, that is to say in a redundant way, by means of different teams using independently developed tools. However, the dynamic use of the configuration database can not be verified in such way because it involves the interaction (or "integration") between the control code and the configuration data. Such important task is automated by the algorithm provided in this paper, such to multiply of several times the temporal advantages mentioned above for a single instantiation of the abstract test-suite. If executed manually, this activity would require many hundreds of person-hours and is also subject to the risk of human errors (i.e. omission of a certain test condition), as witnessed by past experiences.

---

[7] Actually, the algorithm ensures the coverage of "at least" all the conditions provided in the railway condition table, but usually much more test conditions are generated.

A successful real application of this approach has been the IXL of the city of Manchester. This system controls a wide area around the city and needs a large configuration database (about 300 Megabytes), considering all the logic conditions required by the control tables. In order to manage such a complexity, a multistage installation was needed, starting from a basic configuration (including only routes of a certain area) and extending it step-by-step until the ultimate configuration (including more than 100 routes). In order to verify the system, about 1500 abstract tests have been defined and their instantiation produced more than 200,000 executable tests. Each stage of the installation involved several versions of the application software. Since the whole project was based on 5 stages, each one involving an average of 10 software versions, the test activity would have been nearly unfeasible without the abstract testing approach presented in this paper.

## 5. Conclusions

This paper has presented an abstract testing methodology suitable for a class of control systems. The methodology allows for an automatic instantiation of functional tests in order to achieve an extensive configuration coverage in a reasonable time. The application of the approach to railway interlocking installations allowed test engineers to verify such systems in a both effective and efficient way: effectiveness consisted in a more extensive detection of configuration-related errors, while efficiency was represented by the significant gain in time and effort with respect to traditional approaches. The control systems of several large European railway stations are significant success stories for the approach.

The abstract testing approach described in this paper could be applied to a range of critical applications in different fields, when the software architecture of the control systems fits or can be engineered/refactored to fit the reference models presented in this paper. Applications of interest include other types of railway and automotive control installations, including traffic management and brake-by-wire systems; supervisory control systems; flexible manufacturing installations; etc. The required customizations consist in implementing the abstract testing algorithm in a proper framework, allowing interfacing with the abstract testing representation formalism, configuration database and test scripting environments, according to the specific simulation languages.

Future refinements of the methodology will focus on the development of a general model-based testing framework, supporting a stronger integration between the state-based specification formalism, the configuration data and the simulation environments. At the state, in fact, the process could be significantly improved from the cohesion viewpoint: the available tools have been developed separately and their interaction must be often provided by hand, with an avoidable waste of time. The objective of generality also regards the standardization of the interfaces, with XML[8]- based representations possibly playing an important role. As an example, this would enable the possibility for a railway authority to apply the same test specification to different IXL suppliers, by asking them to develop interfaces (or adaptation layers) which are compliant to the defined standards (for both the representation of configuration data and the commands, stimuli and outputs of simulation environments).

---

[8] The eXtended Markup Language (XML) is widely used as a versatile language for data representation and exchange [30].

## Disclaimer

In this paper no reserved data about products or technologies owned by Ansaldo Segnalamento Ferroviario S.p.A. have been used.

## References

1. CENELEC EN50126 Railways Applications: The specification and demonstration of Reliability, Maintainability and Safety, 1999
2. Telelogic AB: Telelogic Tau Logiscope 6.1 Test Checker - Basic Concepts, 2004
3. Model-Based Testing: http://www.model-based-testing.org
4. Holcombe, M.: An integrated methodology for the specification, verification and testing of systems. In: Software Testing, Verification and Reliability, Wiley, Volume 3, Issue 4, November 1993, pp. 149–163
5. De Nicola, G.; di Tommaso, P.; Esposito, R.; Flammini, F.; Marmo, P. & Orazzo, A.: A Grey-Box Approach to the Functional Testing of Complex Automatic Train Protection Systems. In: LNCS Vol. 3463, Springer-Verlag, The Fifth European Dependable Computing Conference, EDCC-5, Budapest, Hungary, April 20–22, 2005, pp. 305–317
6. Flammini, F.; di Tommaso, P.; Lazzaro, A.; Pellecchia, R. & Sanseviero, A.: The Simulation of Anomalies in the Functional Testing of the ERTMS/ETCS Trackside System. In: Proceedings of the 9th IEEE Symposium on High Assurance Systems Engineering, HASE'05, Heidelberg, Germany, October 12–14, 2005: pp. 131–139
7. Clarke, James M.: Automated Test Generation from a Behavioral Model. In: Proceedings of Pacific Northwest Software Quality Conference, May 1998
8. Hartman, A.; Kirshin A. & Nagin K.: A Test Execution Environment Running Abstract Tests for Distributed Software. In: Software Engineering and Applications (ed. M. H. Hamza), SEA 2002, Cambridge, USA, November 4–6, 2002, pp. 448–453
9. OMG Unified Modeling Language: http://www.omg.org/uml
10. OMG Model Driven Engineering: http://www.omg.org/mde
11. Bondavalli, A.; Dal Cin, M.; Latella, D.; Majzik, I.; Pataricza, A. & Savoia, G.: Dependability analysis in the early phases of UML-based system design. In: Computer Systems Science and Engineering, CRL Publishing Ltd, Vol. 16, Issue 5, 2001, pp. 265–275
12. Abbaneo, C.; Flammini, F.; Lazzaro, A.; Marmo, P.; Mazzocca, N. & Sanseviero, A.: UML Based Reverse Engineering for the Verification of Railway Control Logics. In: Proceedings of Dependability of Computer Systems, DepCoS'06, Szklarska Poręba, Poland, May 25–27, 2006, pp. 3–10
13. Pickin, S.; Jard, C. ; Le Traon, Y.; Jéron, T. ; Jézéquel, J.M. & Alain Le Guennec. System Test Synthesis from UML Models of Distributed Software. In: Proceedings of the 22nd IFIP WG 6.1 International Conference on Formal Techniques for Networked and

Automatic instantiation of abstract tests on specific configurations for large critical control systems      21

Distributed Systems, LNCS vol. 2529, Springer-Verlag, FORTE'02, Houston, Texas, November 2002, pp. 97–113
14. Offutt, J.; Liu, S.; Abdurazik, A. & Ammann, P.: Generating test data from state-based specifications. In: Software Testing, Verification and Reliability, Wiley, Volume 13, Issue 1, March 2003, pp. 25–53
15. Clarke, E.M. & Wing, J.M.: Formal Methods: State of the Art and Future Directions. In: ACM Computing Surveys, Vol. 28, Issue 4, 1996, pp. 626–643
16. Brave Y. & Heymann, M.: Control of discrete event systems modeled as hierarchical state machines. In: IEEE Transactions on Automatic Control, Volume 38, Issue 12, 1993, pp. 1803–1819
17. Chen, Y. L. & Lin, F.: Modeling of discrete event systems using finite state machines with parameters. In: Proceedings of the IEEE International Conference on Control Applications, 2000, pp. 941–946
18. Statestep: http://www.statestep.com/
19. Senesi, F. & Malangone, R.: Formal Method Analysis and Evaluation of ERTMS Test Specification for the Italian High-Speed Railway. In: Proceedings of the 6th FORMS/FORMAT Symposium (ed. E. Schnieder, G. Tarnai), Braunschweig, Germany, January 25–26, 2007: pp. 97–106
20. Müllerburg, M.: Systematic testing: A means for validating reactive systems. In: Software Testing, Verification and Reliability, Wiley, Volume 5, Issue 3, October 2006, pp. 163–179
21. Bogdanov, K. & Holcombe, M.: Statechart testing method for aircraft control systems. In: Software Testing, Verification and Reliability (Wiley), Volume 11, Issue 1, March 2001, pp. 39–54
22. Ibrahim, K.E.F. & Whittaker, J.A.: Model-based Software Testing. In: Encyclopedia on Software Engineering (ed. J. J. Marciniak), Wiley, 2001
23. Myers, G.J.: The Art of Software Testing, 2nd Edition. Wiley, 2004
24. Cimatti, A.; Giunchiglia, F.; Mongardi, G.; Romano, D.; Torielli, F. & Traverso, P.: Formal Verification of a Railway Interlocking System using Model Checking. In: Formal Aspects of Computing, Springer-Verlag, Volume 10, Issue 4, 1998: pp. 361–380
25. Havelund, K.; Lowry, M. & Penix, J.: Formal Analysis of a Space-Craft Controller Using SPIN. In: IEEE Transactions on Software Engineering, Volume 27, Issue 8, August 2001, pp. 749–765
26. UNISIG: ERTMS/ETCS Class 1 Issue 2.2.2 Subset 026, 2002
27. Amendola, A.M.; Impagliazzo, L.; Marmo, P. & Poli, F.: Experimental evaluation of computer-based railway control systems. In: Proceedings of the 27th IEEE International Symposium on Fault-Tolerant Computing, FTCS'97, 1997, pp 380–384
28. di Tommaso, P.; Esposito, R.; Marmo, P. & Orazzo, A.: Hazard Analysis of Complex Distributed Railway Systems. In: Proceedings of the 22nd IEEE International Symposium on Reliable Distributed Systems, SRDS'03, 2003: pp. 283–293
29. Heath, W.S.: Real-Time Software Techniques. Van Nostrand Reinhold, New York, 1991
30. Benz, B & Durant, J.R.: The XML Programming Bible, 2nd edition. Wiley Publishing Inc., New York, 2003
31. Hwang, J.G. & Lee, J.W.: Laboratory integration testing of railway signalling systems for high-speed trains. In Computers in Railways IX, WIT Press, 2004

| VARIABLE | BRIEF DESCRIPTION |
|---|---|
| TC | Test-case of the abstract test specification |
| I_States | Subset of input states defined by the abstract test |
| Input_States | Domain of possible input states for the system under test |
| Si | Condition used to extract a subset from the set of input states |
| STATEi | Generic input state amongst the ones defined by the abstract test |
| Input_Sequence | Sequence of inputs needed to reach a certain state |
| Input_Sensors | Subset of sensors involved in the abstract test |
| Sensor_List | Complete set of sensors available in the installation under test (configuration data) |
| SEN | Condition used to extract a subset from the set of sensors |
| Sensor | Generic sensor amongst the ones involved in the abstract test |
| Input_TC | Input of the test case, selected from the complete input domain (it specializes to each sensor to which it is applied) |
| Input | Complete input domain for the system under test |
| I | Condition used to extract a subset from the input domain |
| INPUTj | Generic input amongst the ones defined by the abstract test |
| Output_Actuators | Subset of actuators involved in the abstract test |
| Actuator_List | Complete set of actuators available in the installation under test (configuration data) |
| ACT | Condition used to extract a subset from the set of actuators |
| Actuator | Generic actuator amongst the ones involved in the abstract test |
| O | Output condition to be checked on actuators |
| Attributes | Set of attributes defining output state as defined in the abstract test |
| STATEo | Output state defined in the abstract test, intended in the algorithm as the subset of attributes to be checked |
| Attrib | Generic attribute whose value has to be checked for correctness |
| S_Attrib | Attribute of sensor processes which has the same name of the one to be checked |
| Sen_LP | Set of logic processes associated to sensors and containing at least one of the attributes to be checked |
| Sensor_Association_List | Set of logic processes associated to sensors |
| A_Attrib | Attribute of actuator processes which has the same name of the one to be checked |
| Act_LP | Set of logic processes associated to actuators and containing at least one of the attributes to be checked |
| Actuator_Association_List | Set of logic processes associated to actuators |
| Logic_Processes | Set of logic processes associated to sensors or actuators and containing at least one of the attributes to be checked |
| Logic_Process | Generic logic process associated to sensors or actuators and containing at least one of the attributes to be checked |
| LP_Attrib | Attribute of logic processes associated to sensors or actuators which has the same name of the one to be checked |
| Proc_Attrib | Complete set of system attributes having the same name of the one to be checked |
| So | Condition to be checked on attributes of system output state |

Table 1. Variables used in the abstract testing algorithm.